\documentclass[preprint]{aastex}

\begin{document}


\shorttitle{WFI~J2026$-$4536 and WFI~J2033$-$4723:\\ Two New Quadruple Gravitational Lenses\altaffilmark{1}}
\title{WFI~J2026$-$4536 and WFI~J2033$-$4723:\\ Two New Quadruple Gravitational Lenses\altaffilmark{1}}






\author{Nicholas~D.~Morgan\altaffilmark{2,3},
        John~A.~R.~Caldwell\altaffilmark{4,5},
	Paul~L.~Schechter\altaffilmark{3},
	Alan~Dressler\altaffilmark{6},
	Eiichi~Egami\altaffilmark{7},
        Hans-Walter~Rix\altaffilmark{5}
}



\altaffiltext{1}{Based on observations obtained with the MPG/ESO 2.2~m telescope, the CTIO 1.5~m telescope, and the Clay and Baade 6.5~m telescopes at the Las Campanas Magellan Observatories.  Also based on observations with the NASA/ESA Hubble Space Telescope, obtained at the Space Telescope Science Institute, which is operated by the Association of Universities for Research in Astronomy, Inc., under NASA contract NAS 5-26555.  These observations are associated with HST program \#9744.}

\altaffiltext{2}{Center for Astronomy \& Astrophysics, Yale University, New Haven CT 06520; morgan@astro.yale.edu}

\altaffiltext{3}{Department of Physics, Massachusetts Institute of Technology, Cambridge MA 02139; schech@space.mit.edu}

\altaffiltext{4}{Space Telescope Science Institute, 3700 San Martin Drive, Baltimore MD 21218; caldwell@stsci.edu}

\altaffiltext{5}{Max-Planck-Institut f\"ur Astronomie, K\"onigstuhl 17, Heidelberg, Germany; rix@mpia-hd.mpg.de}

\altaffiltext{6}{The Observatories of the Carnegie Institution of Washington, 813 Santa Barbara St., Pasadena, CA 91101; dressler@ociw.edu}

\altaffiltext{7}{Steward Observatory, University of Arizona, Tucson, AZ 85721; eegami@as.arizona.edu}


\begin{abstract}
We report the discovery of two new gravitationally lensed quasars, WFI~J2026$-$4536 
and WFI~J2033$-$4723, at respective source redshifts of $z=2.23$ and $z=1.66$.  Both 
systems are quadruply imaged and have similar PG1115-like image configurations.  
WFI~J2026$-$4536 has a maximum image separation of 1\farcs4, a total brightness of 
$g = 16.5$, and a relatively simple lensing environment, while WFI~J2033$-$4723 has a 
maximum image separation of 2\farcs5, an estimated total brightness of $g \approx 17.9$, and a more 
complicated environment of at least six galaxies within 20\arcsec.  The primary lensing 
galaxies are detected for both systems after PSF subtraction.  Several of the broadband flux ratios
in these systems show a strong ($0.1-0.4$ mags) trend with wavelength, 
suggesting either microlensing or differential extinction through the lensing galaxy.  
For WFI~J2026$-$4536, the total quasar flux has dimmed by 0.1 mag in the blue but only 
half as much in the red over three months, suggestive of microlensing-induced variations.  
For WFI~J2033$-$4723, resolved spectra of some of the quasar components reveal emission line flux 
ratios that agree better with the macromodel predictions than either the broadband or 
continuum ratios, also indicative of microlensing.  The predicted differential time 
delays for WFI~J2026$-$4536 are short, ranging from 1-2 weeks for the long delay, but 
are longer for WFI~J2033$-$4723, ranging from 1-2 months.  Both systems hold promise for future
monitoring campaigns aimed at microlensing or time delay studies.

\end{abstract}




\section{Introduction}

Gravitationally lensed quasars are useful laboratories for a number of cosmological studies, 
such as measuring the Hubble constant \cite{koop01}, estimating the surface density of 
intermediate-redshift galaxies \cite{koch02}, and constraining the composition and structure 
of lensing dark matter halos \cite{sche02, dala02}.  In many of these studies, particularly 
those focusing on the properties of the lensing galaxies, quadruply imaged quasars are more 
valuable than their doubly imaged counterparts because they probe several lines of 
sight through the lens and provide additional constraints when constructing potential models.
From this perspective, the relatively high number of quads among the inventory of known lenses 
(1 out of every 3-4 systems) is a welcome abundance.

In this paper, we report the discovery of two new quads in the Southern hemisphere, the 
lensed quasars WFI~J2026$-$4536 and WFI~J2033$-$4723.  These systems are the first two 
gravitationally lensed quasars discovered as part of an optical survey for lenses using 
the MPG/ESO 2.2~m telescope operated by the European Southern Observatory (ESO) at La Silla, 
Chile.  A brief outline of the survey along with the discovery observations for the two systems 
are presented in \S\ref{ssd} below.  Follow-up optical and IR imaging of the two quads is 
presented in \S\ref{imaging2026} (WFI~J2026$-$4536) and \S\ref{imaging2033} (WFI~J2033$-$4723), 
and we describe spectroscopic observations of both systems in \S\ref{spec}.  The data leave 
no doubt that the systems are gravitationally lensed quasars, and we comment on archival 
multi-wavelength properties in \S\ref{multiwav}, present models of the lensing potentials in 
\S\ref{models}, and discuss the evidence for microlensing in \S\ref{fluxratios}.  Finally, 
we summarize our findings and outline possible directions for future work in 
\S\ref{conclusions}.  Unless noted otherwise, we assume $H_{\rm{o}} = 75$~km~s$^{-1}$~Mpc$^{-1}$ 
and an ($\Omega_{m},\Omega_{\Lambda}$)~=~(0.3,0.7) universe.

\section{Survey Overview and ESO 2.2m Discovery\label{ssd}} 

The two systems reported here were discovered from a wide-field imaging survey for lensed quasars 
in the Southern hemisphere using the MPG/ESO 2.2~m telescope.  Our strategy is to 
identify gravitational lens candidates by searching for objects that appear multiply imaged on scales
of $\sim$1\arcsec\ and that also possess quasar-like colors using standard UVX color selection.  
In contrast to traditional ``targeted'' optical surveys for lenses, 
we do not re-image known quasars but instead search a wide region of sky for any objects that meet our 
morphology and color criteria.  This affords two advantages over searches built on existing 
quasar catalogs.  First, a multiply imaged quasar can appear as a single, extended source in the mediocre 
seeing (2\arcsec-3\arcsec\ FWHM) characteristic of Schmidt photographic plates (Kochanek 1991).  Such lenses 
may evade quasar-finding algorithms programmed to find point-like objects, and may never make it into
quasar catalogs culled from the Schmidt-telescope surveys of the last two decades.  Second, the pool of bright 
quasars that have fueled previous targeted lensed quasar surveys in the South is mostly exhausted.  The 
Calan-Tololo (Maza et al. 1996) and Hamburg-ESO (Wisotzki et al. 2000) quasar surveys have accounted for 
eight of the nine bright (total flux of $I\lesssim18$) lensed quasars discovered south of 
$\delta=-10\degr$.\footnote{See the CfA-Arizona Space Telescope Lens Survey (CASTLES) at http://cfa-www.harvard.edu/castles.}
The large majority of bright quasars from these surveys have already been re-imaged as part
of targeted lens surveys from the ground (Wisotzki et al. 1993, 1996, 1999, 2003; 
Claeskens et al. 1996; Morgan et al. 1999) and with the {\it Hubble Space Telescope} (HST; Gregg et al. 2001; 
Morgan et al. 2003).  Thus the bulk of new, optically-bright lensed quasars in the South will likely 
have to come from outside of current quasar catalogs.

The wide-field survey described here consists of $UBR$ broadband images obtained with the 
Wide-Field Imager (WFI) CCD, an 8k$\times$8k mosaic camera on the MPG/ESO 2.2~m telescope with a field 
of view of 33\arcmin\ $\times$ 34\arcmin.  We obtain $UBR$ images at each telescope
pointing and essentially treat each triplet of images as a free-standing dataset.  Approximately 
1000 square degrees of overlapping $UBR$ data have been obtained to date.  Exposure times are 
typically 60 s in $U$ and 30 s in $B$ and $R$, which corresponds to a limiting magnitude of $B\approx22$.
Our survey areas lie outside the Galactic plane ($|b|>30\degr$) and mostly in the South Galactic Cap 
between $\delta = -20\degr$ to $-60\degr$ and $\alpha = 20^{\rm{h}}$ to $6^{\rm{h}}$, and
also a smaller region in the North Galactic Cap south of $\delta = -5\degr$.  The fairly-good seeing 
from La Silla (median $R$-band FWHM of 1\farcs0 for our survey data) and CCD sampling 
(0\farcs24 pixel$^{-1}$) facilitate the identification of double and quadruple image configurations 
with separations as small as 0\farcs7.  Roughly speaking, since there is $\sim$1 bright ($B \lesssim 18$) 
quasar per square degree (Boyle et al. 1988), and the lensing probability is $\sim1\%$ for $B\sim18$ 
magnitude quasars (Kochanek 1991), we expect to find on the order of 10 bright gravitationally 
lensed quasars.

All survey data are processed using a customized pipeline.  Source detections and 
instrumental magnitudes are obtained using a variant of DoPHOT (Schechter, Mateo \& 
Saha 1993), modified to use an empirical point spread function (PSF) for point source
photometry.  Each WFI image typically contains between 500-1000 objects, which is a 
sufficient number such that color zeropoints can be obtained from the instrumental 
colors alone.  For example, the $B-R$ colors for each mosaic are calibrated using 
the ``backbone'' of the $B$ vs. $B-R$ color-magnitude diagram (e.g., Caldwell \& 
Schechter 1996, their Figure~5), which corresponds to the main sequence turnoff for 
high Galactic latitude stars.  The $U-B$ colors are calibrated by aligning
the instrumental main sequence in color-color space to the calibrated stellar main 
sequence for each mosaic.  This yields a well-calibrated 
color-color diagram for each WFI pointing and allows us to separate quasar candidates 
from main-sequence stars using a straightforward color cut.  This is illustrated in 
Figure~1, which shows a sample $U-B$ vs. $B-R$ color-color plot of about 300 square 
degrees (970 mosaic pointings) of survey data.  The contours trace the stellar main 
sequence, and the solid curve is a synthetic quasar track as a function of
redshift computed using the FIRST Bright Quasar Survey (FBQS) composite quasar spectrum 
(Brotherton et al. 2001).  Objects that can be decomposed into multiple point sources with 
at least one component's colors or the group's ensemble colors blueward (below) the dashed 
line in Figure~1 are flagged as gravitational lens candidates.

Both WFI~J2026$-$4536 (20$^{\scriptsize{\mbox{h}}}$ 26$^{\scriptsize\mbox{m}}$ 
10$\fs$43, $-45\arcdeg$ 36$\arcmin$ 27$\farcs$1; J2000.0; hereafter J2026) and 
WFI~J2033$-$4723 (20$^{\scriptsize{\mbox{h}}}$ 33$^{\scriptsize\mbox{m}}$ 42$\fs$08, 
$-47\arcdeg$ 23$\arcmin$ 43$\farcs$0; J2000.0; hereafter J2033) were identified as 
candidate gravitational lenses from the above procedure.  The original 
survey data for J2026 were obtained over a two-week period during April and May 2002,
and the data for J2033 were obtained during September 2001.  Postage stamps of the 
$R$-band discovery images for both systems are shown in Figure~2 (insets).  DoPHOT 
resolved both targets into three components in the $B$ and $R$ filters (components A,
B and C in Figure~2) and two components (A and B) in the $U$ filter.  The AB image 
separation is 1\farcs4 for J2026 and 2\farcs5 for J2033.  Both systems were also 
flagged as color-selected quasar candidates (see Figure~1).  For J2026, the internally calibrated 
($U-B$,~$B-R$) colors of components A and B were (-0.7,~0.7) and (-0.6,~0.5), 
respectively.  For J2033, bluer colors of (-1.0,~0.7) and (-0.9,~0.7) were obtained.  
The multiple image morphology and quasar-like colors provided the initial suspicion that 
the objects might be gravitationally lensed quasars, and motivated the follow-up 
observations discussed below.

Figure~2 shows 1\arcmin$\times$1\arcmin\ finder charts centered on both systems.  The environment for 
J2026 (Figure~2a) is relatively clean, with only one other prominent galaxy (G1) in the nearby 
field.  In contrast, the environment for J2033 (Figure~2b) is densely populated, with at least 6 
galaxies (G1-G6) within 20\arcsec, and several other relatively faint objects (X1-X4) whose 
profiles are point-like but have yet to be spectroscopically identified.

\section{WFI~J2026$-$4536 -- Follow-up Imaging and Analysis\label{imaging2026}}

Follow-up observations of J2026 were obtained on several occasions, with the primary 
goals of resolving the system's image morphology, judging the amount of quasar 
variability, and searching for the foreground lensing galaxy.  The quadruple nature 
of the system was first confirmed during optical re-imaging with the Magellan 6.5~m 
Clay telescope in April 2003.  The lensing galaxy was first detected from IR imaging 
with the Magellan 6.5~m Baade telescope in September 2003.  Second epoch optical data were 
obtained with the Clay telescope in August, providing a handle on the system variability
over a three-month baseline.  The system was also imaged with HST in October 2003 
yielding precise relative astrometry between the system components 
and a magnitude estimate of the lensing galaxy.  We describe each of these observations 
below, beginning with the HST data.

\subsection{HST/NICMOS -- Image Morphology and the Lensing Galaxy\label{hstwfi2026}}

We observed J2026 with the HST/NICMOS NIC2 camera (plate scale of 0\farcs07565 pixel$^{-1}$)
on 21 October 2003 under the Cycle 12 HST Imaging of Gravitational Lenses program 
(PI:~C.~Kochanek; PID 9744).  Observations consisted of four dithered {\tt MULTIACCUM} 
exposures through the F160W filter ($\approx$ {\it H}-band) for a total on-source integration 
of about 46 minutes.  After standard {\tt CALNICA} processing, bad pixel masks were created 
for each dithered exposure using the drizzle software of Fruchter \& Hook (1997), which 
were used during the multi-component fitting described below.  A drizzled image of the four 
exposures is shown in Figure~3a and clearly reveals a quadruple image morphology.  One can 
identify the three components detected in the WFI discovery image as the unresolved A1 and A2 
components, along with components B and C to the North and West respectively.  The system 
configuration, particularly the bright and closely-separated A1 and A2 images, is a classic 
inclined-quad lensing configuration (see Saha \& Williams 2003), similar to the Northern hemisphere 
quad PG~1115+080 \cite{weym80}, and leaves no doubt that the system is a gravitational lens.

For each dithered image, we have modeled the system's light distribution using four PSFs 
generated with the TinyTim v6.1a software of Krist \& Hook (2003).  The PSFs were generated 
on a 10$\times$10 oversampled grid to allow for accurate subpixel shifts, taking into 
account the system's position on the NICMOS chip and the PAM focus position.  Minimization
was performed using a Powell (Press et al. 1992) routine.  Figure~3b shows the drizzled 
residuals after subtracting the best-fit models.  A fifth component is clearly
revealed interior to the four point sources.  The object has a peak countrate of $\sim$1\% 
of component A1 and is extended with respect to the NICMOS PSF (gaussian FWHM of 0\farcs27 
compared to 0\farcs13 for point sources).  Its position and extent readily identify it as 
the foreground lensing galaxy (hereafter component G).  

We repeated the fit including a circularly-symmetric deVaucoulers profile (convolved 
with the NICMOS PSF) to account for the lensing galaxy, and solved for the relative positions
and intensities of the five components along with the effective radius of the galaxy profile.
The averages of the best-fit parameters among the four exposures are listed in Table~1.  
The separation between the brightest two components is small, only 0\farcs33, while the 
largest separation (A1-B) is 1\farcs44.  The average effective radius of the galaxy model 
was 0\farcs51 $\pm$ 0\farcs18, where the error is from the standard deviation among the four 
frames.  

To compute the galaxy magnitude, we first used an aperture radius of 
0\farcs38 (5 NICMOS pixels) to sum the galaxy flux in the residual images after the 
four-component subtraction.  The galaxy light in this region is clearly discernible from 
background and is mostly unaffected by the subtraction residuals, yielding $H=19.56\pm0.05$.
We then computed an aperture correction to large radii ($\sim$11 $R_{\mbox{eff}}$) of 
1.13 magnitudes using a synthetic deVaucouleur's profile (convolved with the PSF) defined by 
the average $R_{\mbox{eff}}$ of the lensing galaxy.  The largest source of systematic error 
for the final galaxy magnitude given in Table~1 is the effective radius used to compute the aperture 
correction.  Varying $R_{\mbox{eff}}$ by 50\% changes the final galaxy magnitude by 
$\pm$0.4 mags, which is not included in the rms error listed in Table~1.  Figure~3c shows 
the drizzled residuals after subtracting the best-fit five component model, and shows 
a relatively clean subtraction apart from residuals in the A1 and A2 cores.

\subsection{Magellan 6.5~m -- Optical Variability\label{imaging2026mag}}

To examine the system's variability, we obtained first and second epoch observations 
of J2026 on 24 and 25 April 2003 and 1 and 4 August 2003 at the Las Campanas Magellan Observatories.  
Data were obtained using the Clay telescope equipped with the Magellan Instant Camera (MagIC), 
a 2k$\times$2k CCD array with a plate scale of 0\farcs06926 pixel$^{-1}$.  A summary of the 
observations is given in Table~2.

The April data were bias-subtracted, sky-flattened, and trimmed using standard procedures.  
Stacked subrasters of the $i$-band exposures are shown at different contrast levels in Figures~3d and 3e.  The same 
image morphology is seen as from the HST data, although the A1 and A2 images are only marginally 
separated in $i$-band.  For the $ugr$-band images, components A1 and A2 appear as a single 
source because of the poorer seeing.  

To obtain the broadband photometric properties of the system, we simultaneously fit four 
empirical PSFs to each Magellan frame.  For the $i$-band data, we allow the position and 
flux of all four components to vary, but we fix the relative positions for the remaining 
frames at the averaged $i$-band results.  The final relative quasar positions agreed with the NICMOS
values to within the rms scatter among the $i$-band frames, and the final $ugri$ quasar 
magnitudes are listed in Table~1.  The proximity of A1 to A2 gives rise to a $\sim6\%$ 
scatter in their fluxes, but the sum is tightly constrained with a scatter in $i$ of only 2 
millimag.  Because of the poorer seeing in the $ugr$ data, only the combined A1 and A2 
fluxes are given for these filters.  The C/B ratio is constant to within 0.02 mag across $ugri$ 
and agrees with the NICMOS F160W value, but the (A1+A2)/B ratio shows a significant increase 
(becoming less equal) of 0.35 mag from $u$ to $i$ and 0.44 mag from $u$ to F160W.  Components 
B and C are also systematically bluer than the combined A1+A2 flux by about 0.2, 0.1 and 0.05
mags across $u-g$, $g-r$, and $r-i$ colors, respectively.  This trend is consistent with a color-dependent 
dimming of either A1 or A2.  Unfortunately, the Magellan exposures are not deep enough to 
provide broadband colors of the lensing galaxy, as seen by the absence of significant 
structure in the stacked $i$-band residuals (Figure~3f).

For photometric calibration, the Magellan photometry was placed onto the standard Sloan 
magnitude system using zeropoints derived from the Landolt (1992) standard SA 110-232 and 
the synthetic Johnson-Kron-Cousins/Sloan magnitude conversions of Fukugita et al. (1996).  
To help calibrate future observations, Table~1 also lists relative photometry obtained 
from empirical PSF fitting for eight field stars within 2\arcmin\ of the lens.

The August $ugri$ data were all taken in non-photometric conditions.  The seeing was 
worse than for the April observations, ranging from 1\farcs3 to 1\farcs9 in $ugr$ and 
0\farcs7 to 1\farcs0 in $i$, and the four components were resolvable in only two of the 
$i$-band frames.  To measure the total system variability, we have modeled the light 
profile using fixed relative offsets from the April $i$-band data and calibrated the fluxes 
using the local standards in Table~1.  The total flux from the local standards, measured 
with respect to each filter's PSF star, agreed with the April measurements to better than 
0.005 mag in $ugi$ and to $0.016$ mag in $r$.  We found that the total flux from J2026 has dimmed 
with respect to the April values by 0.10, 0.08, 0.06, and 0.05 magnitudes in $ugri$, 
respectively.  

\subsection{Magellan 6.5~m -- IR Field Standards}

The lensing galaxy for J2026 was first detected from IR imaging with the Magellan Baade telescope on 
14 September 2003 using the Persson's Auxiliary Nasmyth Infrared Camera (PANIC),
which has a 2.2\arcmin\ square field of view and a plate scale of 0\farcs1267 pixel$^{-1}$.  
Dithered 15 second observations were obtained through $H$ and $Ks$ filters with average 
seeing of 0\farcs34 FWHM.  The cumulative exposure times were 14 minutes in $H$ and 18 
minutes in $Ks$.  The stacked $Ks$-band images of J2026 are shown in Figures~3g and 3h, and
the corresponding residuals after subtracting a four-component empirical PSF model are shown in Figure~3i.
Although the galaxy is visible in the residual panel, the quality of the fit is somewhat poor 
with significant systematic residuals from each quasar subtraction arising from PSF variations 
across the chip.  The residuals from the over-subtracted cores and under-subtracted wings are 
stronger than the peak galaxy signal in both the $H$ and $Ks$ images, which makes it difficult to derive an 
accurate galaxy color.  However, the PANIC observations do give 
a sense for the quality of observations that can be expected from future ground-based IR monitoring
of this system.  To facilitate detecting future IR variability, we give $H$- and $Ks$-band magnitudes 
of the four quasar components and nearby field stars intended for use as local standards in Table~1.  
Magnitudes are again from empirical PSF fitting, with the PSF star placed onto the standard system 
using Persson et al. (1998) standards \#9115 and \#9172.  

\section{WFI~J2033$-$4723 --  Follow-up Imaging and Analysis \label{imaging2033}}

The Magellan follow-up observations for J2033 proceeded along similar lines as for J2026.  We 
obtained follow-up images with the Clay telescope on 1 August 2003 using the same instrumental 
setup described in \S\ref{imaging2026mag}.  Single exposures in $ugr$ and several exposures in 
$i$ were obtained, all in non-photometric conditions.  The seeing FWHM ranged between 
1\farcs0 to 1\farcs3 in $ugr$ and between 0\farcs5 and 0\farcs6 in $i$.  A summary of the 
observations is given in Table~2.

The data were reduced in the same manner as for J2026.  Figure~4a shows the stacked $i$-band 
image of J2033 and clearly shows four point sources, with component A from the discovery 
image resolved into components A1 and A2.  The image geometry is another inclined-quad, 
similar to J2026 and again conclusively identifying it as a gravitational lens.  To search for 
the lensing galaxy, we simultaneously fit and subtracted four empirical PSFs to each $i$-band 
frame.  The averaged residuals from the six frames are shown in Figure~4b.  A fifth object 
is now visible in the residual panel, with a peak flux of $\sim$3\% that of A1.  
The object is also extended with a gaussian FWHM of 0\farcs81 compared to 0\farcs56 for 
point sources, making its identification as the lensing galaxy certain (hereafter component 
G).  We attempted to model the galaxy with a circularly symmetric deVaucouleurs profile, but the 
effective radius was poorly constrained.  Instead, we use a circularly symmetric gaussian 
profile and solve for the positions and relative fluxes of the five components for each of the
$i$-band frames.  The model results are listed in Table~3.  To estimate the lensing galaxy 
magnitude, we used a large (4\farcs1) aperture radius on the stacked residuals for the 
four-component models.  No significant residual structure remains after the five-component 
subtraction, as shown by the stacked residuals in Figure~4c.

Components A1 and A2 were marginally separable by eye in $r$, but appeared as one source in 
the $u$ and $g$ images because of the poorer seeing.  To model the $ugr$ data, we fixed the 
relative separations at the average $i$-band results and solved for the relative fluxes and 
the system's overall position.  The galaxy flux in these filters is negligible, so only the 
four-component model was used.  The solutions are again listed in Table~3.  As seen for J2026, 
the flux ratios vary with filter, with the C/B ratio systematically increasing (becoming more 
equal) by 0.23 mags from $u$ to $i$ and the (A1+A2)/B ratio systematically increasing 
(becoming less equal) by 0.10 mag from $u$ to $i$.  In $r$, where the A1 and A2 fluxes are 
separable, the (A1+A2)/B change is mostly attributable to a 0.05 mag drop in component A2's 
flux from $i$ to $r$. 

For future photometric reference, Table~3 also presents photometry from empirical PSF fitting 
for 7 field stars within 2\arcmin\ of J2033.  The magnitudes have been calibrated using 
the Landolt (1992) standard Feige~22.  Since conditions were nonphotometric during the night, 
there may be unknown systematic errors in the magnitudes.  The $i$-band PSF star had an rms 
scatter in its aperture magnitude of 0.09 mags over the course of the six consecutive $i$-band 
exposures, so zeropoint uncertainties of $\sim0.1$ mag may be present in each filter.
Regardless, since magnitudes were determined from relative PSF fitting, the field stars will 
provide a local set of standards to gauge future variability.  The colors from Table~3 are 
consistent with the known stellar and quasar locus in the Sloan filter system (e.g., Smith et al. 2002, 
Richards et al. 2001), so zeropoint errors are mostly grey.

\section{Spectroscopy\label{spec}}

\subsection{CTIO 1.5~m Observations\label{ctiospec}}

To identify the quasar redshifts, we obtained unresolved spectral observations of both systems 
on 31 July 2003 using the R-C spectrograph on the CTIO 1.5~m telescope.  The low-dispersion 
grating \#13 was used, providing an effective wavelength coverage of 3600~\AA\ to 8000~\AA\ and
a dispersion of 5.73~\AA\ pixel$^{-1}$.  Three 20 minute exposures of each target were 
taken through a 2\arcsec\ slit width.  Several comparison exposures of He-Ar lamps were 
also obtained for wavelength calibration.  The spectra for each object were reduced using 
standard IRAF routines and flux calibrated using the spectrophotometric standard 
Fiege~110.

The unresolved spectra of the two systems are shown in Figures~5a and 5b.  Both show typical 
quasar broad emission lines and underlying continuum.  For J2026, a gaussian fit to the peak 
(top 25\% of flux) of the CIV line gives an emission redshift of $z=2.223 \pm 0.001$.  The CIV
feature in quasar spectra is typically blueshifted with respect to the systemic redshift, and 
indeed we do find a higher redshift estimate of $z=2.237 \pm 0.001$ from the Ly$\alpha$, CII 
and OI/SiIII blended emission lines, but the shift is an order of magnitude larger than 
that typically found for quasars \cite{vand01}.  For J2033, we find a systemic redshift of 
$z=1.661 \pm 0.003$ based on a gaussian fit to the peak of the MgII profile.

\subsection{Magellan 6.5~m Observations\label{magellanspec}}

We also obtained resolved spectral observations of the four quasar images for J2033 on 
15~September 2003 using the Inamori Magellan Areal Camera and Spectrograph (IMACS; 
Bigelow \& Dressler 2003) on the 
Baade telescope.  The resolved spectra are valuable for studying effects that depend on 
position in the lensing galaxy, such as from differential extinction or microlensing.  
The spectrograph was operated in the long camera mode and read out with 2$\times$2 binning, 
providing a final pixel scale of 0\farcs22 pixel$^{-1}$.  The 300~lines~mm$^{-1}$ grating was 
used, yielding a dispersion of 1.5~\AA\ for each binned pixel and an effective wavelength 
coverage of 3800 to 8500~\AA.  We obtained a 300 second exposure of components A1 and A2
and a 600 second exposure of components B and C with the 0\farcs75 wide slit held parallel to the respective
components in each integration.  Exposures of He-Ar lamp lines were also obtained for 
wavelength calibration.

The spectra were bias-subtracted and flattened using standard procedures.  As can be seen in 
the insets to Figure~6, components B and C are well-resolved along the spatial direction but 
components A1 and A2 overlap by a fair amount.  We used an iterative routine to disentangle 
the components during the extraction.  Each dispersion row was modeled
as two overlapping gaussian profiles plus a constant offset for the background sky.  For the 
B and C spectra, the centers, FWHMs, and heights of the gaussians were allowed to vary as a 
function of wavelength on the initial pass but the two centers and FWHMs were replaced by 
smoothly varying polynomials as a function of wavelength on subsequent passes.  On the final 
pass, only the heights of the profiles were free to vary.  The procedure was similar for the 
A1 and A2 spectra except that their relative separation was fixed using the Magellan 
astrometry and the IMACS plate scale derived from the B \& C exposure.  The best-fit gaussian 
FWHMs from the extraction ranged from 0\farcs5 to 0\farcs7, systematically decreasing toward
the red.  The extracted spectra were then flux calibrated using observations of the 
spectrophotometric standard LTT7987 \cite{hamu94}.  The extracted and flux-calibrated spectra 
for the four components are shown in Figure~6.  The gap at 6100 \AA\ is from the physical gap 
between CCDs in the IMACS camera.

As expected, the individual spectra reveal similarly-redshifted quasars consistent with the 
CTIO spectrum from \S\ref{ctiospec}, with appropriately redshifted CIV, CIII], and MgII 
emission lines for each component.  We are primarily interested in the component flux ratios 
for the broad emission lines and the surrounding continuum for what follows.  The line fluxes
were determined by fitting a line to the underlying continuum using $\Delta \lambda = 200$ \AA\ 
windows bracketing each emission profile and then summing the continuum-subtracted flux inside
a $\Delta \lambda = 200$ \AA\ window centered on each profile.  
Table~4 lists the resulting line fluxes and corresponding wavelength windows.  The 
reported errors bars are based on the uncertainty in the underlying continuum, computed by 
shifting the continuum up and down by its zeropoint uncertainty.  We compare the emission 
line flux ratios to the broadband values and model predictions in \S\ref{fluxratios} below.

\section{Multi-Wavelength Properties\label{multiwav}}

The multi-wavelength properties of the systems were explored by querying the NASA/IPAC 
Extragalactic Database around each target's coordinates.  No objects were uncovered within 
2\arcmin\ from J2033, but the search did identify a coincident X-ray source from the ROSAT 
Bright Source Catalog \cite{voge99} within 20\arcsec\ from J2026.  The offset is consistent 
with the 1-2 $\sigma$ positional accuracy of the catalog, so the match is plausible.  The 
quoted ROSAT count rate is 6.8$\pm$2.1 $\times$ 10$^{-2}$ cts s$^{-1}$ and is several times
brighter than other lensed quasars also detected by ROSAT (e.g., PG~1115+080,  
1.9$\pm$0.8 10$^{-2}$ cts s$^{-1}$;  RXJ~0911+0551, 2.0$\pm$0.9 10$^{-2}$ cts s$^{-1}$;
Voges et al. 2000).

\section{The Lensing Potentials\label{models}}

\subsection{WFI~J2026$-$4536\label{2026model}}

The inclined-quad configuration for J2026 implies that the lensed source is located close 
the diamond caustic but away from the caustic cusps, which produces the two bright and nearly 
merging components straddling the outer critical curve.  If we zero our coordinate system at 
the lensing galaxy, then the shear axis should point somewhere between the two merging 
components and the outer critical image.  This is roughly East-West for J2026 and toward
galaxy G1, the only other prominent galaxy in the field.

We modeled the image positions using the fiducial singular isothermal sphere embedded in an 
external shear (model ISx).  This model has an effective projected potential given by 
\begin{equation}
\phi(r, \theta) = br - \frac{\gamma}{2}r^2\cos2(\theta - \theta_{\gamma}),
\end{equation}
where $b$ is the potential strength (measured in arcseconds), $\gamma$ is the shear strength, 
and the adopted sign convection is for $\theta_{\gamma}$ to point toward (or away) from the 
perturbing mass responsible for the shear.  We used the F160W relative positions from Table~1 
as constraints and solved for the strengths of the potential and shear, the shear orientation,
and the galaxy position using the {\tt gravlens} software of Keeton (2003).  The measured 
galaxy position was not used as a constraint (denoted by model ISx+).  Along with the source position, 
this gave seven model parameters for the eight position constraints.  The best-fit model
gave $b=0\farcs6536$, $\gamma=0.120$ and $\theta_{\gamma}= -90\fdg5$ East of North.  The 
galaxy center was placed at ($\Delta\alpha$, $\Delta\delta$) = (0\farcs0753, -0\farcs8168) with respect 
to image B, which is between 2-3 standard deviations from its measured position.  Overall, the
model-predicted and observed quasar positions differed by an rms of 0\farcs0096.  Even though 
this is formally some five times larger than the scatter in the measured quasar positions, 
the accuracy is fairly good when compared to other quad lenses with comparable astrometric 
precision and ISx potential models, which can give rms residuals in the 0\farcs02-0\farcs04 range.  As can be seen in 
Figure~7a, the source position is indeed close to the inner caustic and gives rise to a net 
magnification over the unlensed source of about a factor of 30.

The gravitational time delay $\tau$ as a function of the image position $\vec{\theta}$ is 
given by
\begin{equation}
\tau = \frac{1+z_l}{c^2}\frac{D_l D_s}{D_{ls}}\frac{\left[\frac{1}{2}\left(\vec{\theta}-\vec{\beta}\right)^2 - \phi\right]}{206 265},
\end{equation}
where $\vec{\beta}$ is the source position and $D_l, D_s$, and $D_{ls}$ are angular diameter
distances between the observer and lens, observer and source, and lens and source, 
respectively.  Since the lens redshift for J2026 is unknown, we give time delay estimates 
using the quantity in square brackets which can be computed using the observed image 
positions and ISx model parameters alone.  In Table~5 we give the predicted image positions for 
the four images and lensing galaxy, the image magnifications and the time delays.  A negative 
magnification denotes a parity flip in the lensed image, arising from a saddlepoint in the time 
delay surface.  Images A2 and C are saddlepoints, which can also be seen from the time delay 
contours in Figure~7b.  The leading image for the lens is B, followed by A1, A2, and C.

Lacking a spectroscopic redshift for the lensing galaxy, we can estimate its redshift using 
the potential model and the IR magnitude of the lensing galaxy from \S\ref{hstwfi2026}.  The 
potential strength $b$ is related to the halo velocity dispersion $\sigma$ by 
\begin{equation}
b = \frac{D_{ls}}{D_{s}}\frac{4\pi \sigma^2}{c^2}
\end{equation}
(Narayan \& Bartlemann 1999).  Knowing $\sigma$, the galaxy's $B$-band luminosity can be 
estimated from the Faber-Jackson (1976) relationship $L/L_{\star} = \left(\sigma/\sigma_{\star}\right)^{\gamma}$, 
where $L_{\star}$ corresponds to a $B$-band absolute magnitude of 
$M_{B,\star} = -19.7 + 5\log h$ (assuming $H_{\rm{o}} = 100h$ km s$^{-1}$ Mpc$^{-1}$) and we adopt $\gamma=4.0$ and $\sigma_{\star}=220$ 
km s$^{-1}$ appropriate for elliptical galaxies.  Given $M_B$, the galaxy's observed F160W 
magnitude then follows from 
\begin{equation}
m_{F160W} = M_B + DM(z_l) + K_{B,F160W}(z_l),
\end{equation}
where $DM$ is the cosmological distance modulus and $K_{B,F160W}$ is the generalized 
K-correction from the galaxy's rest-frame $B$-band magnitude to the observed-frame F160W 
magnitude (e.g., Hogg et al. 2002).  The K-corrections are computed using the Bruzual \& 
Charlot (2003) spectral evolution code.  We use a solar-metallicity, passively-evolving 
instantaneous burst model where all stars form at $z_f = 3.0$, which is consistent with the 
observed colors of most other lensing galaxies (Kochanek et al. 2000).  Figure~8a shows the 
predicted $m_{F160W}$ magnitudes as a function of the lens redshift (curved solid line) and 
the observed F160W galaxy magnitude from \S\ref{hstwfi2026} (horizontal line).  In principle, 
the matching redshifts are around $z_l\sim0.4$ and $z_l\sim1.9$, although the higher redshift is ruled out 
since a velocity dispersion of 520 km s$^{-1}$ would be required to produce 
the observed image separations.  At $z_l=0.4$, the required lens velocity dispersion is
reasonable, 180 km s$^{-1}$ (0.4 $L_{\star}$), and the ISx model predicts a 
long time delay (B-C) of 7.5 days and a short delay (A1-A2) of 2.4 hours.  Placing the lens 
at $z_l = 0.3$ or $z_l = 0.6$ yields B-C delays of 5.3 or 12.8 days, respectively.

The shear PA points within 8\degr\ of galaxy G1 and identifies it as the likely source of the 
perturbation.  At a distance of 7\farcs4 from the position of the lensing galaxy, the shear 
strength implies a potential strength for G1 of $b = 1\farcs78$, or 2.7 times larger than for 
the primary lens.  This compares well with the relative fluxes of the two galaxies.  The galaxy's
potential strength $b$ should scale with the square root of its luminosity for a $\gamma=4$ 
Faber-Jackson exponent, so we expect a G/G1 flux ratio of 7.4 based on the ISx+ model.  The 
observed F160W flux ratio inside a 0\farcs38 aperture radius is 6.0, roughly consistent with the
prediction.  

\subsection{WFI~J2033$-$4723}

The image configuration for J2033 is qualitatively similar to J2026, so we started with 
the ISx model with the lens galaxy first fixed at the observed $i$-band position.  This 
gave five parameters for the eight position constraints.  The best-fit model yielded an rms between 
predicted and observed positions of 0\farcs085, almost an order of magnitude worse compared 
to the ISx+ model for J2026.  Allowing the center of the potential to float (model ISx+) improved 
the situation somewhat, with a final rms of 0\farcs056, but is still poor with respect to the J2026
model results.

One way to improve the model is to allow more angular freedom in the potential.  Figure~2b 
shows that two shear axes may be needed: the galaxy G1 due West of the lens and the 
G4-G5-G6 and G2-G3-X3 groups to the North and South.  We again adopted a ISx model for the main
lensing galaxy but also placed a SIS potential at the position of G1 (model ISISx).  The 
position of both the lensing galaxy and G1 were fixed at their measured positions in Table~3, 
giving six parameters for the eight position constraints.  This gave a final rms between 
observed and predicted positions of 0\farcs029 and is roughly a factor of two improvement over the 
ISx+ model with only one additional free parameter.  The potential strength of G1 is larger 
than for the main lensing galaxy, $b=1\farcs24$ compared to $b=0\farcs94$, and the shear 
strength is $\gamma = 0.253$ and points at a PA of 10\fdg5 East of North.  This roughly 
coincides with the location of the G4-G5-G6 and G2-G3-X3 groups.

One final improvement is to allow the center of the lensing galaxy's potential to float 
(ISISx+).  This leaves no degrees of freedom and we expect a perfect fit to the observations. 
The best-fit model reproduced the observed image positions precisely, but the final lensing 
galaxy position shifted only 0\farcs041 from the average $i$-band position.  This is 
well within our measurement errors for component G.  The potential strength of the lensing galaxy was 
$b=0\farcs98$ and the shear direction was $\theta_{\gamma}$ = 12\fdg5 East of North.  This is
mostly unchanged from the ISISx model, but the strengths of G1 and the shear dropped to $b=0\farcs95$
and $\gamma = 0.225$, suggesting that this may be a more realistic model.

In Table~6 we give the observed and predicted positions for the four images from the ISISx+ 
model, as well as the image magnifications and predicted time delays.  Figure~7c shows the 
source position near the inner edge of the diamond caustic and that the inner caustic is 
shifted by some 2\arcsec\ westward of the lensing galaxy center because of the G1 
perturbation.  The net amplification of the background quasar is about a factor of 20 over 
its unlensed flux.

One can see from the time delay contours in Figure~7d that images A2 and C are saddlepoints 
and images A1 and B are minima.  The time delay order is the same as for J2026, namely B, 
A1, A2, and C.  Figure~8b shows the estimated $i$-band magnitude of the lensing galaxy as a 
function of redshift as constrained by the system's image separation and source redshift, 
using the same method described in \S\ref{2026model}.  The predicted magnitudes suggest a
lens redshift in the vicinity of $z_l\sim0.4$ or $z_l\sim1.5$, but the higher redshift is again ruled 
out by the implied velocity dispersion (990 km s$^{-1}$).  The required 
velocity dispersion at $z_l = 0.4$ is 220 km s$^{-1}$, or roughly an $L_\star$ galaxy.  Assuming a lens 
redshift of $z_l=0.4$, the largest differential time delay (B-C) is 26.0 days and the shortest 
(A1-A2) is 1.0 days, although the long delay ranges from 18.0 to 47.0 days for respective 
lens redshifts of $z_l = 0.3$ or $z_l = 0.6$.

\section{Anomalous Flux Ratios and Evidence for Microlensing\label{fluxratios}}

Several of the observed flux ratios for J2026 and J2033 vary with wavelength and differ 
significantly from the model predictions.  For example, the (A1+A2)/B flux ratios for both 
systems drop by 0.1-0.3 mags from $i$ to $u$, while the C/B $i$-band ratios differ by 
0.2-0.3 mags from the model values.  Both trends are much larger than the photometric accuracy.  Such anomalous
flux ratios are found in other lensed quasars as well, and several theories have emerged to 
explain the differences.  Metcalf \& Zhao (2002) argue that unmodeled substructure near the 
lensing galaxy can significantly perturb the magnifications of the lensed images while 
leaving the image positions mostly unchanged.  This produces a constant achromatic effect 
and essentially implies that the problem stems from the use of oversimplistic macromodels.  
Schechter \& Wambsganss (2002) argue that microlensing by stars in the lensing galaxy can 
significantly perturb the observed flux ratios, particularly by demagnifying the flux 
from saddlepoint images, leading to a time-varying chromatic effect.  Of course, more
mundane explanations like dust extinction are also possible (e.g., Falco et al. 1999) when working with optical
fluxes, leading to a constant chromatic effect that can be difficult to disentangle from 
microlensing-induced variations.

While a combination of these factors might be at work for any given system, microlensing 
is perhaps the easiest to identify because of the time and color dependence.  In particular, 
the microlensing-induced variations ought to depend on the intrinsic source size: larger 
sources will cover more of the caustic network, which will wash out the microlensing signal 
and leave flux ratios that approach the macromodel values.  For quasar accretion disks, this 
leads to stronger and more frequent continuum variations in the blue.  At the other extreme, 
the quasar's broad line region, which is at least an order of magnitude larger than the 
continuum-emitting disk (Kaspi et al. 2000, Kochanek 2003), ought to be much less affected by 
microlensing.  An example of the latter can be found in the quadruple quasar HE~0435-1223 
(Wisotzki et al. 2003), which shows emission line ratios that agree better with the
model predictions than do the broadband flux ratios, although the match is still not perfect.

For J2026, we find evidence for microlensing from the system's broadband flux ratios 
and time variability.  The combined (A1+A2)/B flux ratio shows a strong 
wavelength dependence, systematically increasing from 4.29 in $u$ to 5.90 in $i$, 
a difference of 0.35 mags.  The trend is likely due to varying A1+A2 
flux rather than from component B, since the C/B ratio is constant to within 0.02 mag 
from $u$ to $i$ and thus each component is probably not significantly affected by microlensing or dust.  
This implies that there is less A1+A2 flux at bluer wavelengths
than expected from the overall shape of the B and C quasar spectra.  The likely 
causes are either dust extinction or microlensing demagnification of one of the
A components.  Patchy dust has an even chance of obscuring either A1
or A2, but a significant microlensing demagnification should preferentially occur
in the saddlepoint image A2 according to Schechter \& Wambsganss (2002).  Although 
the poor seeing prevents us from determining which component is responsible 
for the trend, the second epoch observations do suggest that microlensing is present.  The 
three-month variability is larger in the blue ($\Delta m = 0.10$ mag) 
than in the red ($\Delta m = 0.05$ mag), consistent enhanced microlensing fluctuations expected from
smaller source sizes.

It is also interesting that the C/B flux ratio, although mostly constant from $u$ to $Ks$
and thus probably not significantly affected by microlensing or dust, still differs by 0.2 
mags from the model prediction.  In this case, the discrepancy may reflect a shortcoming in 
the model which used only the image positions as constraints.  To check this, we repeated the 
ISx+ model using the observed $i$-band C/B ratio as a constraint, but we had to tighten the B and
C flux uncertainties to 0.1\% before the model successfully reproduced the obseved ratio.  
The final rms between the observed and predicted quasar positions was 0\farcs015, still 
typical of other quadruple lenses, but the required shear increased by 25\% and the final
galaxy position shifted by 0\farcs032, or about 4$\sigma$ from its measured position.  So 
the model can be forced to fit C/B ratio, although the stronger shear and misaligned center 
suggests it is less realistic.  More complicated models, perhaps taking into account
asymmetries or substructure in the galaxy potential, are probably required to give a 
satisfying result.

For J2033, we find evidence for microlensing mostly from the C/B flux ratio.
The ratio increases from 0.75 in $u$ to 0.93 in $i$, a difference of 0.23 mags,
again suggesting either dust extinction or microlensing.  However, even the $i$-band ratio 
differs by 0.3 mags from the model prediction of 1.22.  Yet the agreement 
is strikingly improved when the model is compared to the IMACS emission line flux ratios.  
Figure~9a shows the $m_C - m_B$ magnitude differences obtained from the emission 
line fluxes and bracketing continuum.  Each emission feature has three measurements: 
the line ratio from Table~4 and the flanking continuum ratios from the IMACS spectra.  
The continuum ratios systematically tend toward the model prediction (horizontal line) 
as one moves from blue to red, as do the broadband flux ratios (open circles).  However, 
the emission line ratios are systematically larger than the bracketing continuum 
for each emission feature, and are mostly in agreement with the macromodel predictions.  The 
effect is most noticeable for CIV and CIII], yielding C/B ratios clearly greater than 
unity, a feature not found in either the continuum or broadband ratios.  The different 
emission line and continuum ratios rule out significant dust extinction, which ought to 
affect both measurements equally, leaving microlensing as the likely explanation for the
flux ratio trend.
There is also evidence for 0.2 mag variability in the C/B flux ratio between the Magellan
broadband measurements and the continuum ratios, suggesting 
some combination of microlensing and intrinsic quasar variability over the 1 month baseline.

A smaller color trend is found for the (A1+A2)/B broadband ratios, which increase 
from 2.72 in $u$ to 2.99 in $i$, a 0.10 mag difference.  Figure~9b shows the A2/A1 
IMACS ratios similar to the C/B plot described above, but the noise from 
the extraction masks any trend that might be present in the emission line or continuum ratios.
At best, one can say that the line ratios are consistent with the model predictions and 
broadband flux ratios.

\section{Summary and Conclusions\label{conclusions}}

We have reported the discovery of two bright gravitationally lensed quasars found during a 
wide-field optical survey for lenses in the Southern hemisphere.  WFI~J2026$-$4536 
($g = 16.5$, source redshift of $z_s = 2.23$) and WFI~J2033$-$4723 ($g \approx 17.9$, $z_s = 1.66$) are 
both quadruple systems that share similar inclined-quad image configurations with maximum 
image separations of 1\farcs4 and 2\farcs5, respectively.  The lensing galaxies are detected 
for both systems.

For J2026, the small image separation leads to relatively short differential time delays
ranging from 1-2 weeks for the long delay (B-C) to several hours for the short delay 
(A1-A2), depending on the lensing galaxy's redshift.  Even the long delay will be difficult 
to measure from optical monitoring campaigns unless the quasar is uncharacteristically 
variable on such timescales.  This may limit the usefulness of J2026 for cosmological 
applications but makes it an attractive target for microlensing studies since intrinsic
variability will not be easily confused with microlensing-induced signals.  The (A1+A2)/B
flux ratio varies by 0.37 mags from $u$ to $Ks$, strongly suggesting either microlensing
or dust extinction, while the enhanced variability observed in the blue over three months
suggests at least some microlensing-induced variability.  Ground-based monitoring will help
to gauge the presence of microlensing, but will be a challenge for the A1 and A2 components
given the small image separation.

The ROSAT detection of J2026 also opens the prospect of measuring an X-ray time delay between 
the short-delay components.  The A1-A2 delay is on the order of hours and could be measured
during a single Chandra observations if significant X-ray variability is present.  Since
Chandra would be unable to spatially separate the A1 and A2 components, the approach would require
autocorrelating their unresolved lightcurve as originally suggested by Chartas et al. (2001).

For J2033, the larger image separations yield longer differential time delays and will also 
make ground-based monitoring an easier task.  The longest time delay (B-C) is predicted to 
be 1-2 months, again depending on the lens redshift.  However, the lensing potential is
more complicated than for J2026, with a group of at least six galaxies within 20\arcsec\ that
require two separate perturbation axes for effective modeling.  The complex potential may 
complicate the interpretation of any measured delay.  This system is also an attractive target for 
microlensing studies.  The system's C/B flux ratio varies by 0.23 mags from $u$ to $i$, 
while the C/B emission line ratios agree much better with the macromodel predictions than either 
the continuum or broadband values, as expected from microlensing magnifications that ought 
to depend on the source size.  The emission line ratios suggest that the anomalous C/B flux ratio 
in the blue is caused primarily by microlensing-induced perturbations rather than 
unmodeled substructure in the lensing potential, as might be expected given the ratio's strong 
color trend.

The redshifts of both lensing galaxies are still needed to yield a precise time delay 
prediction, and spectroscopic identification of nearby field objects, particularly the nearby 
G1 galaxies and the galaxy group surrounding J2033, will help to construct more accurate 
lensing models.  In view of the two systems' optical brightness, demonstrated variability,
and evidence for microlensing, it is clear that future monitoring will be a worthwhile
undertaking.

\acknowledgments

We would like to thank Sergio Gonz\'alez for his assistance with the CTIO~1.5m observations,
and Chris Kochanek for placing J2026 onto the HST schedule at short notice.  NDM and PLS wish 
to thank the MPIA in Heidelberg for their hospitality during a number of visits, during which 
this work was initiated.  Part of this research (PLS) was supported by US NSF grant AST02-06010.

\clearpage


 

\makeatletter
\def\jnl@aj{AJ}
\ifx\revtex@jnl\jnl@aj\let\tablebreak=\nl\fi
\makeatother


\begin{deluxetable}{cccccccccc}
\rotate
\tablecaption{Relative Astrometry and Photometry for WFI~J2026$-$4536 and Field
\label{TABLE1}}
\tablenum{1}
\tablewidth{0pt}
\tablehead{  
\colhead {ID} &
\colhead {$\Delta\alpha$ (\arcsec)} &
\colhead {$\Delta\delta$ (\arcsec)} &
\colhead {$u$} &
\colhead {$g$} &
\colhead {$r$} &
\colhead {$i$} &
\colhead {$H$} &
\colhead {$F160W$} &
\colhead {$Ks$}
}
\startdata
   A1 & +0.1621(0014)    & -1.4281(0009)    &  ---   &   ---  &   ---  & 17.109(059) & 15.585 & 15.252(013) & 14.978 \\
   A2 & +0.4149(0014)    & -1.2133(0009)    &  ---   &   ---  &   ---  & 17.363(073) & 16.107 & 15.673(008) & 15.371 \\
A1+A2 &    ---           &    ---           & 17.336 & 16.816 & 16.557 & 16.475(003) & 15.062 & 14.690(008) & 14.404 \\
    B &   $\equiv0$      &  $\equiv0$       & 18.916 & 18.581 & 18.450 & 18.402(012) & 17.054 & 16.709(009) & 16.359 \\
    C & -0.5722(0015)    & -1.0436(0004)    & 19.145 & 18.814 & 18.666 & 18.625(011) & 17.208 & 16.916(010) & 16.559 \\
    G & -0.0813(0031)    & -0.7967(0068)    &  ---   &  ---   &  ---   &    ---      &  ---   & 18.43(05)   &  ---   \\
   G1 & -7.398(026)      & -1.940(006)      & 22.948 & 21.931 & 20.863 & 20.160(011) & 17.493 &         --- & 16.848 \\
      &                  &                  &        &        &        &             &        &             &        \\
    7 &  16.5            &  41.9            & 23.500 & 20.774 & 19.519 & 18.597(050) & ---    &         --- & ---    \\
   10 & -50.8            &  40.8            & 18.336 & 16.947 & 16.463 & 16.311(001) & 15.080 &         --- & 15.094 \\
   12 &  49.9            &  36.3            & 17.145 & 15.462 & 14.850 & 14.644(035) & ---    &         --- & ---    \\
   15 &  -6.0            &  21.9            & 21.088 & 19.510 & 18.921 & 18.656(059) & 17.122 &         --- & 17.161 \\
   18 & -23.1            &   9.8            & 20.517 & 19.628 & 19.332 & 19.215(123) & 18.116 &         --- & 18.266 \\
   19 &  31.0            &   1.8            & 21.084 & 20.145 & 19.753 & 19.562(031) & ---    &         --- & ---    \\
   25 &  41.0            & -36.6            & 21.885 & 19.479 & 18.528 & 18.134(028) & ---    &         --- & ---    \\
   28 &  24.7            & -55.0            & 21.918 & 19.761 & 18.961 & 18.630(075) & ---    &         --- & ---    \\
\enddata
\tablecomments{Relative photometry and astrometry results for WFI~J2026$-$4536 and field.  Error bars, when present, are from the rms scatter among frames.  The $ugri$ data are from the April Magellan run.  The G1 position is measured from the Magellan $i$-band frames.  Relative positions for the quasar components and the lensing galaxy are from the HST/F160W frames.  Galaxy G1 $ugri$ photometry was computed using a circular aperture with radius of 0\farcs7, while the F160W magnitude for component G was computed as described in the text.  For the Magellan data, the psf star was \#10 for the the $r,i,H,Ks$ filters, and \#12 for the $u$,$g$ filters.
}
\end{deluxetable}

\clearpage


 

\makeatletter
\def\jnl@aj{AJ}
\ifx\revtex@jnl\jnl@aj\let\tablebreak=\nl\fi
\makeatother


\begin{deluxetable}{lcrccc}
\tablecaption{Summary of Magellan Optical Observations
\label{TABLE2}}
\tablenum{2}
\tablewidth{0pt}
\tablehead{  
\colhead { Target} &
\colhead { filter } &
\colhead { Date (UT)} &
\colhead { $N_{im}$} &
\colhead { $<$Exp$>$ (s)} &
\colhead { $<$FWHM$>$ (\arcsec) } 
}
\startdata
 WFI~J2026$-$4536 &  $u$  & 24 Apr 03   &  1  &  360 &  1.07 \\
                &  $g$  & 24 Apr 03   &  1  &  120 &  0.98 \\
                &  $r$  & 24 Apr 03   &  1  &  120 &  0.90 \\
                &  $i$  & 25 Apr 03   &  3  &  120 &  0.55 \\
                &  $u$  &  1 Aug 03   &  1  &  240 &  1.87 \\
                &  $g$  &  4 Aug 03   &  1  &  120 &  1.52 \\
                &  $r$  &  4 Aug 03   &  1  &  120 &  1.34 \\
                &  $i$  &  4 Aug 03   &  5  &  120 &  0.86 \\
                &       &                  &     &      &       \\
 WFI~J2033$-$4723 &  $u$  &  1 Aug 03   &  1  &  240 &  0.96 \\
                &  $g$  &  1 Aug 03   &  1  &  120 &  1.31 \\
                &  $r$  &  1 Aug 03   &  1  &  120 &  0.94 \\
                &  $i$  &  1 Aug 03   &  6  &  120 &  0.56 \\
\enddata
\end{deluxetable}

\clearpage


 

\makeatletter
\def\jnl@aj{AJ}
\ifx\revtex@jnl\jnl@aj\let\tablebreak=\nl\fi
\makeatother


\begin{deluxetable}{cccccccc}
\tablecaption{Relative Astrometry and Photometry for WFI~J2033$-$4723 and Field
\label{TABLE3}}
\tablenum{3}
\tablewidth{0pt}
\tablehead{  
\colhead {ID} &
\colhead {$\Delta\alpha$ (\arcsec)} &
\colhead {$\Delta\delta$ (\arcsec)} &
\colhead {$u$} &
\colhead {$g$} &
\colhead {$r$} &
\colhead {$i$}
}
\startdata
   A1 &  -2.193(03) &   1.258(02)  &    ---  &   ---  & 18.839 & 18.682(005) \\
   A2 &  -1.477(03) &   1.368(02)  &    ---  &   ---  & 19.346 & 19.144(005) \\
A1+A2 &    ---      &    ---       &  18.749 & 18.473 & 18.311 & 18.136(004) \\
    B &  $\equiv$0  & $\equiv0$    &  19.836 & 19.606 & 19.478 & 19.327(006) \\
    C &  -2.108(03) &  -0.282(03)  &  20.151 & 19.760 & 19.600 & 19.409(012) \\
    G &  -1.412(33) &   0.277(20)  &    ---  &   ---  &   ---  &   19.713    \\
   G1 &  -5.397(10) &   0.247(10)  &    ---  &   ---  &   ---  &    ---      \\
      &             &              &         &        &        &             \\
   32 &  -5.7       &  69.7        & 20.715 & 19.659 & 19.221 & 19.130(006)  \\
   33 &  57.9       &  69.2        & 20.386 & 18.256 & 17.482 & 17.214(001)  \\
   36 & -32.5       &  62.1        & 20.495 & 19.588 & 19.195 & 19.127(005)  \\
   38 &  -0.9       &  61.6        & 19.625 & 17.169 & 16.211 & 15.844(004)  \\
   55 & -15.7       &  15.6        & 17.726 & 16.613 & 16.220 & 16.149(003)  \\
   61 &  60.4       &  -2.4        &  ---   & 19.252 & 18.210 & 17.884(000)  \\
   76 & -40.8       & -43.5        &  ---   & 21.425 & 20.163 & 18.871(003)  \\
\enddata
\tablecomments{Relative photometry and astrometry results for WFI~J2033$-$4723 and 
field.  Error bars, when present, are from the rms scatter among frames.  The psf 
stars were \#55, 38, and 61 for the $u$, $g$, and $r,i$ filters, respectively.  
Galaxy photometry for component G was computed as described in the text.  Note that
the data were obtained under nonphotometric conditions, with possible zeropoint errors
of $\sim0.1$ mag in each filter.
}

\end{deluxetable}

\clearpage


 

\makeatletter
\def\jnl@aj{AJ}
\ifx\revtex@jnl\jnl@aj\let\tablebreak=\nl\fi
\makeatother


\begin{deluxetable}{cccccc}
\tablecaption{Emission Line Fluxes for WFI~J2033$-$4723
\label{TABLE4}}
\tablenum{4}
\tablewidth{0pt}
\tablehead{  
\colhead {Feature} &
\colhead {$\Delta \lambda$ (\AA)} &
\colhead {A1} &
\colhead {A2} &
\colhead {B} &
\colhead {C}
}
\startdata
CIV   & [4000,4200] & 63.3(23.6) & 47.2(21.0) & 49.1(11.5) & 72.5(13.0) \\
CIII] & [4950,5150] & 43.8(8.1)  & 33.0(7.0)  & 27.6(3.7)  & 39.7(3.9)  \\
MgII  & [7350,7550] & 36.2(7.9)  & 22.5(6.2)  & 23.0(3.7)  & 29.1(3.4)  \\
\enddata
\tablecomments{$\Delta\lambda$ defines the wavelength window used to measure 
each emission line flux.  Fluxes are in units of 10$^{-16}$ ergs s$^{-1}$ 
cm$^{-2}$.}
\end{deluxetable}

\clearpage


 

\makeatletter
\def\jnl@aj{AJ}
\ifx\revtex@jnl\jnl@aj\let\tablebreak=\nl\fi
\makeatother


\begin{deluxetable}{cccccccc}
\tablecaption{ISx+ Model Results for WFI~J2026$-$4536
\label{TABLE5}}
\tablenum{5}
\tablewidth{0pt}
\tablehead{  
\colhead {ID} &
\colhead {$\Delta\alpha$ (\arcsec)} &
\colhead {$\Delta\delta$ (\arcsec)} &
\colhead {$\mu$} &
\colhead {$t$} &
}
\startdata
   A1 &  0.1653   & -1.4295   &  12.27 &  -0.1944 \\
   A2 &  0.4069   & -1.2033   & -10.27 &  -0.1927 \\
    B &  0.0091   &  0.0010   &   3.53 &  -0.2956 \\
    C & -0.5764   & -1.0533   &  -3.50 &  -0.1627 \\
    G &  0.0753   & -0.8168   &   ---  &    ---   \\
\enddata
\tablecomments{Listed are the predicted image and galaxy positions, the model magnifications ($\mu$), and dimensionless time delays ($t$) as described in the text.  The coordinate system is centered on the observed position of image B.
}
\end{deluxetable}

\clearpage


 

\makeatletter
\def\jnl@aj{AJ}
\ifx\revtex@jnl\jnl@aj\let\tablebreak=\nl\fi
\makeatother


\begin{deluxetable}{cccccccc}
\tablecaption{ISISx+ Model Results for WFI~J2033$-$4723
\label{TABLE6}}
\tablenum{6}
\tablewidth{0pt}
\tablehead{  
\colhead {ID} &
\colhead {$\Delta\alpha$ (\arcsec)} &
\colhead {$\Delta\delta$ (\arcsec)} &
\colhead {$\mu$} &
\colhead {$t$} &
}
\startdata
   A1 & -2.192    &  1.257     &   7.70 &  -3.997 \\
   A2 & -1.476    &  1.366     &  -5.90 &  -3.981 \\
    B &  0.000    &  0.000     &   3.42 &  -4.274 \\
    C & -2.103    & -0.278     &  -4.18 &  -3.846 \\
    G & -1.419    &  0.326     &   ---  &   ---   \\
\enddata
\tablecomments{Listed are the predicted image and galaxy positions, the model magnifications ($\mu$), and dimensionless time delays ($t$) as described in the text.  The coordinate system is centered on the observed position of image B.
}
\end{deluxetable}

\clearpage


\thispagestyle{empty}
\begin{figure}[h]
\vspace{7.0 truein}
\includegraphics{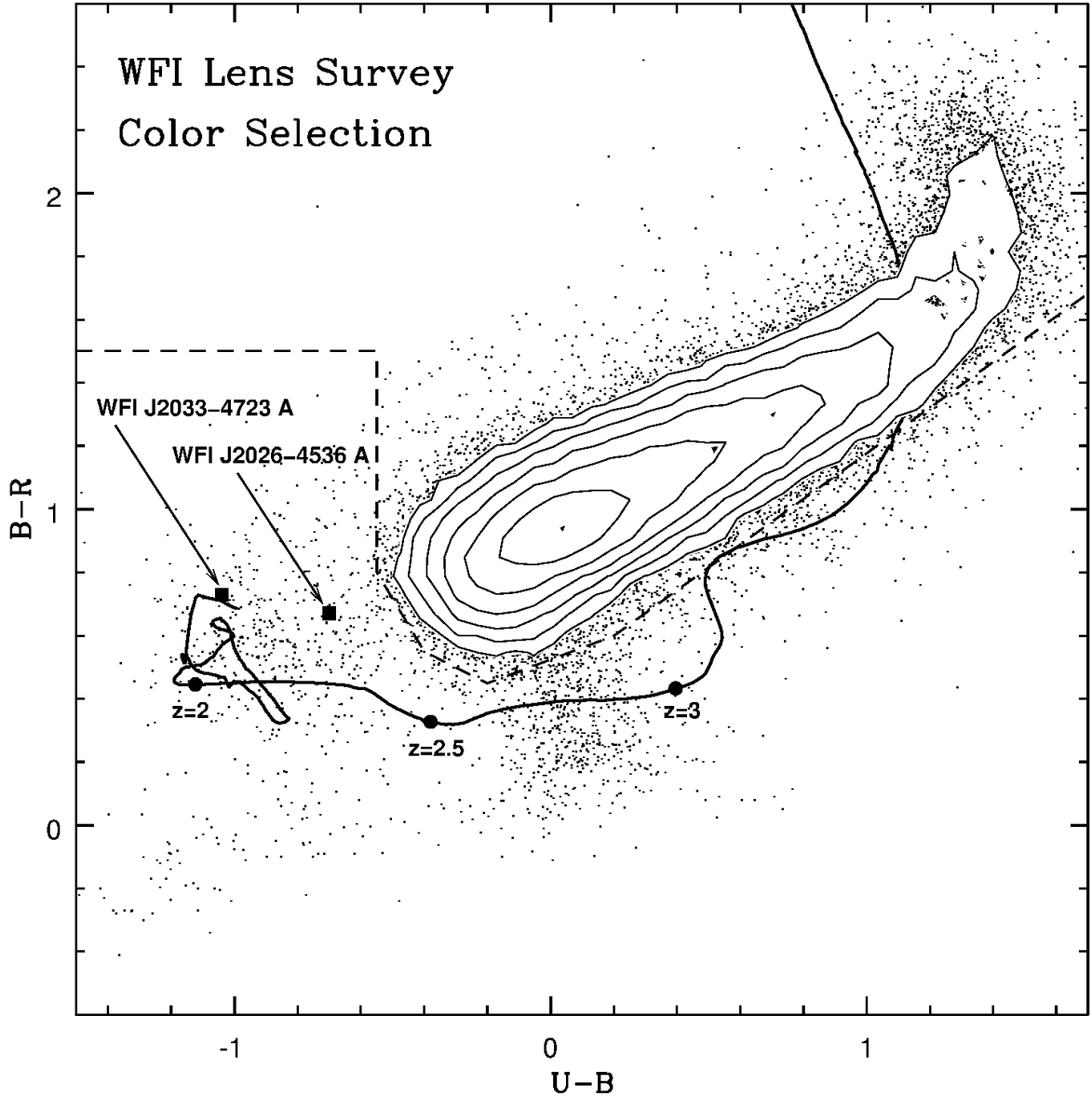}
\caption{Color-color diagram for $\sim$300 square 
degrees of WFI survey data.  Contours trace the stellar main sequence, and the curved solid
line is the quasar color-color track as a function of redshift determined using the FBQS 
composite quasar spectrum.  The dashed line is the color boundary used when selecting 
gravitationally lensed quasar candidates as described in the text.  The positions of the 
combined A1+A2 components (denoted simply by A) for WFI~J2026$-$4536 and WFI~J2033$-$4723 are 
marked by the black squares.}
\end{figure}

\clearpage

\thispagestyle{empty}
\begin{figure}[h]
\vspace{7.0 truein}
\includegraphics{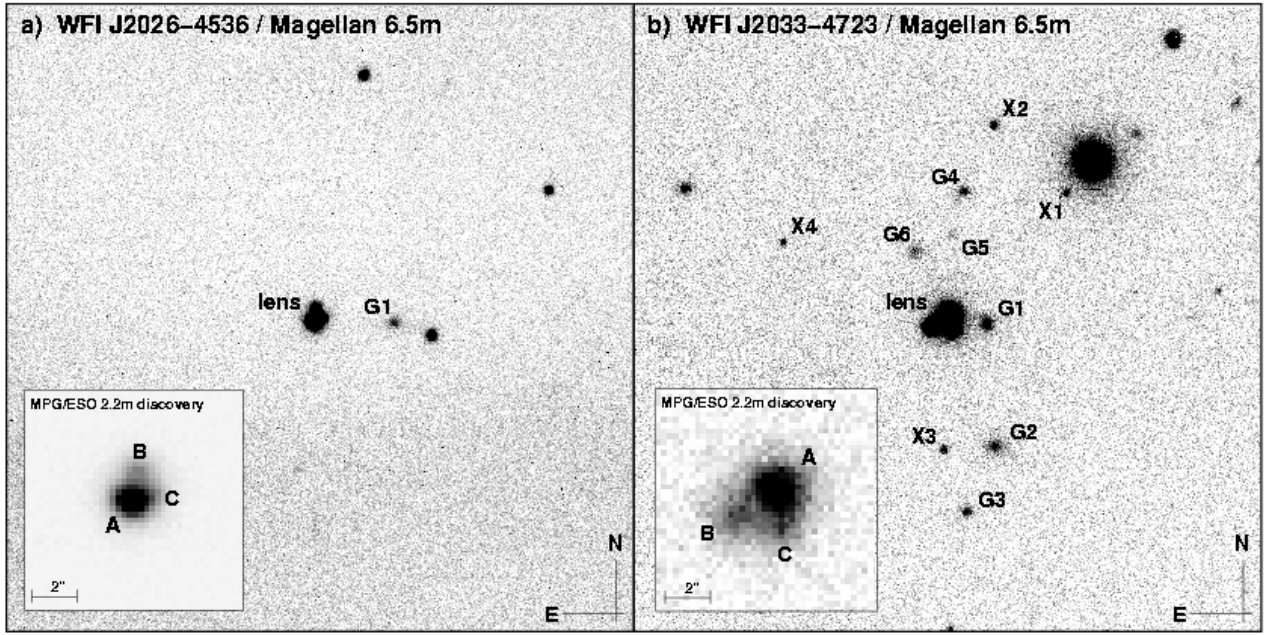}
\caption{{\it Panel a}: Stacked $i$-band observations of 
WFI~J2026$-$4536 taken with the Clay 6.5~m.  The lensed quasar and nearby galaxy (G1) are 
labeled.   {\it Panel b}: Stacked $i$-band observations of WFI~J2033$-$4723 taken with 
the Clay 6.5~m.  Several nearby galaxies (G1-G6) and point-like objects (X1-X4) are labeled.  
Chart sizes are 1\arcmin\ square.  Insets for both panels show the $R$-band discovery images 
from the MPG/ESO 2.2~m.}
\end{figure}

\clearpage

\thispagestyle{empty}
\begin{figure}[h]
\vspace{7.0 truein}
\includegraphics{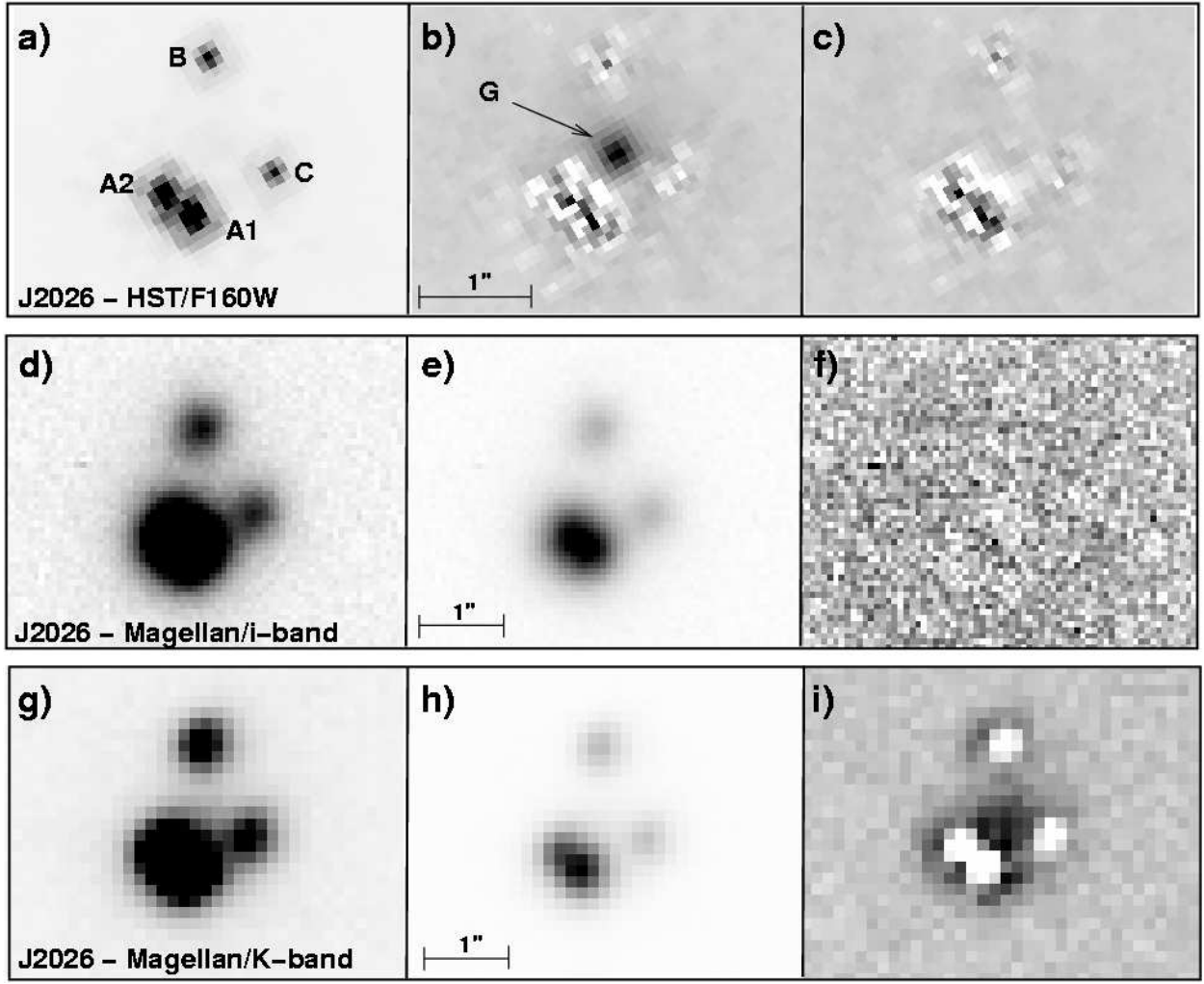}
\caption{{\it Panel a}: Drizzled HST/F160W image of WFI~J2026$-$4536.
{\it Panel b}: Same as (a), but after subtracting the best four-component PSF model.  
{\it Panel c}: Same as (a), but after subtracting the best five-component model including 
a circularly-symmetric deVaucouleurs profile for the lensing galaxy.  The contrast in panels 
(b) and (c) is the same, stretching from $\pm$1 count second$^{-1}$ or equivalently $\pm$2\% of A1's 
peak.  {\it Panel d}: Stacked Magellan $i$-band image of WFI~J2026$-$4536 taken in 0\farcs55 
FWHM seeing.  {\it Panel e}: Same as (d), but at a higher contrast to reveal the A1 and A2 
components.  {\it Panel f}: Same as (d), but after subtracting the best four-component PSF 
model.  {\it Panels g-i}: Same as (d)-(f), but for the stacked Magellan $Ks$-band images.
All panels have North up and East left.}
\end{figure}

\clearpage

\thispagestyle{empty}
\begin{figure}[h]
\vspace{7.0 truein}
\includegraphics{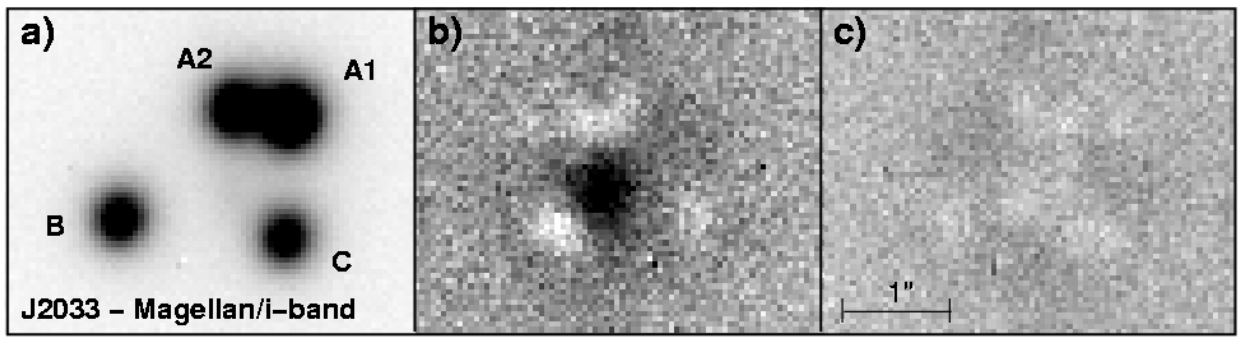}
\caption{{\it Panel a:} Stacked Magellan $i$-band image of 
WFI~J2033$-$4723.  {\it Panel b}: Same as (a), but after subtracting the best four-component 
PSF model.  {\it Panel c}: Same as (a), but after subtracting the best five-component model
including a circularly symmetric gaussian profile for the lensing galaxy.  The contrast for 
panels (b) and (c) stretch from -2$\sigma$ to +5$\sigma$, where $\sigma$ is the expected rms 
per pixel from Poisson and readnoise statistics.  All panels have North up and East left.}
\end{figure}
\clearpage

\thispagestyle{empty}
\begin{figure}[h]
\vspace{7.0 truein}
\includegraphics{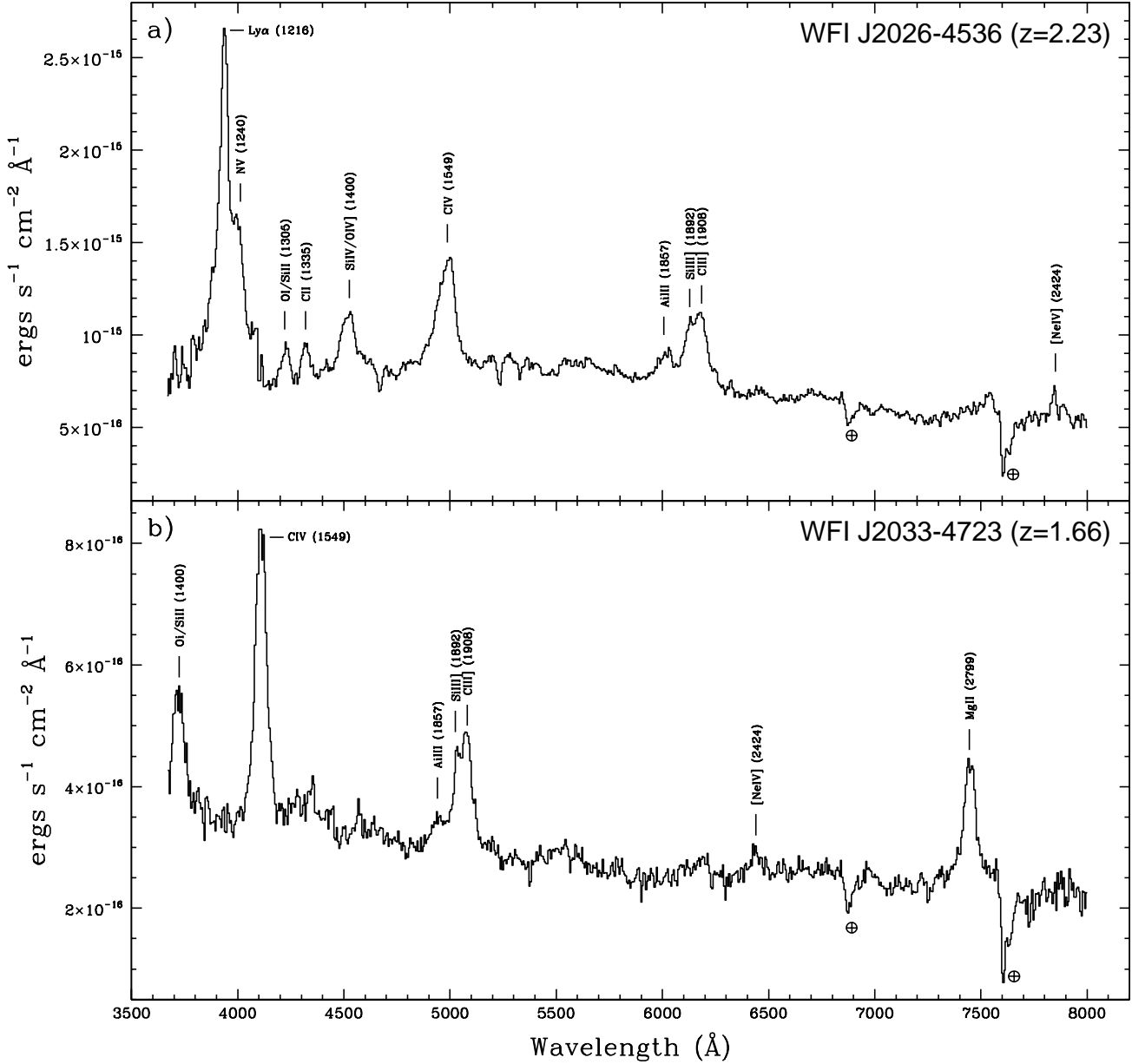}
\caption{Unresolved long-slit spectra of WFI~J2026$-$4536 
({\it panel a}) and WFI~J2033$-$4723 ({\it panel b}) obtained with the CTIO 1.5~m.  Prominent 
emission lines are labeled for both quasars.}
\end{figure}
\clearpage

\thispagestyle{empty}
\begin{figure}[h]
\vspace{7.0 truein}
\includegraphics{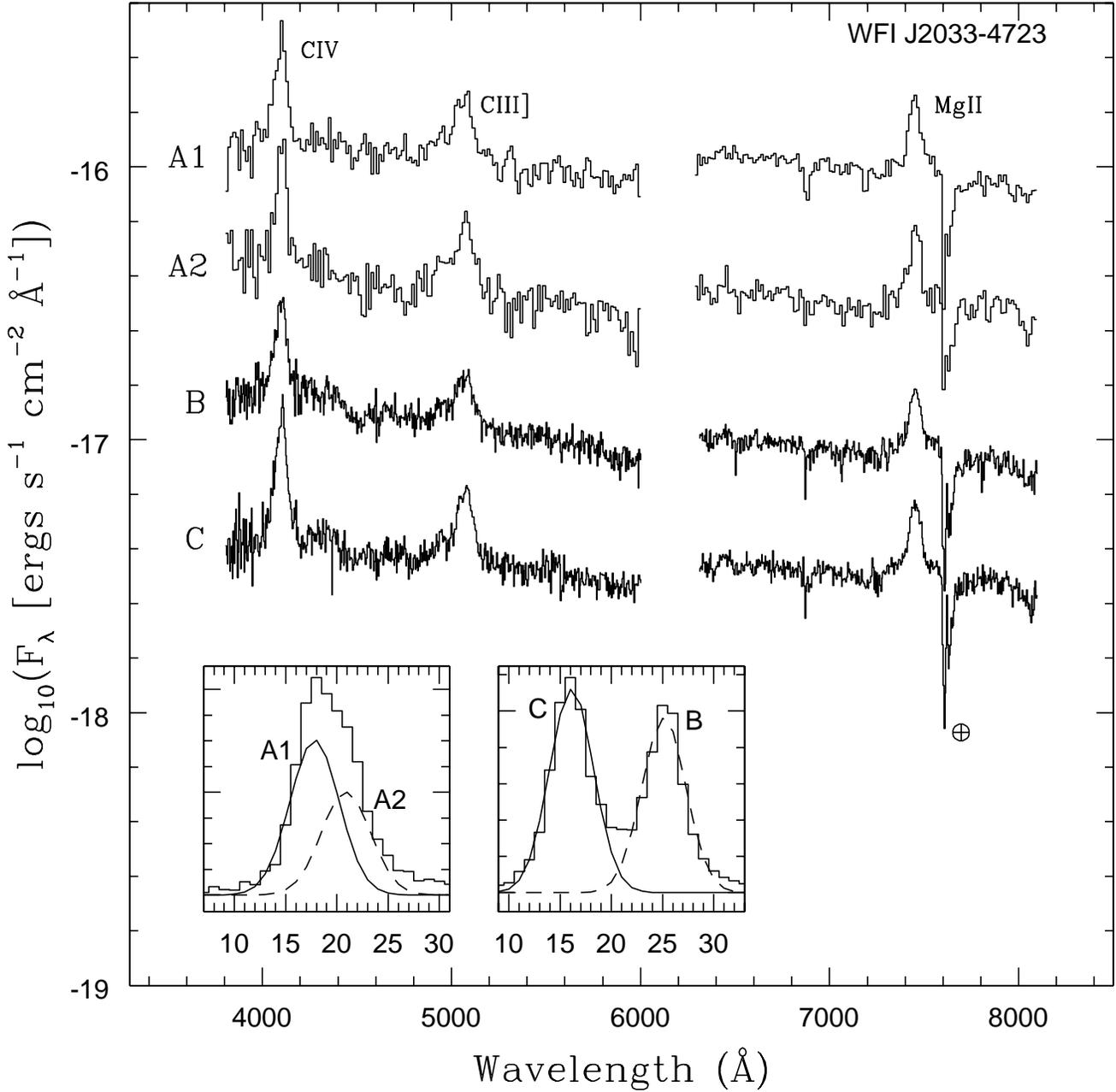}
\caption{Resolved spectra of the four WFI~J2033$-$4723 quasar 
components obtained with the Magellan/IMACS spectrograph.  For clarity, the A1, B, and C 
spectra have been shifted by 0.3, -0.6, and -1.1 dex, respectively.  Bin sizes are 13.5 
\AA\ (9 pixels) for the A1 and A2 spectra and 7.5 \AA\ (5 pixels) for the B and C spectra.  
Insets show a slice along the dispersion direction (as a function of pixel number)
in the vicinity of the MgII emission profile for A1 and A2 (left histogram) and for B and C 
(right histogram), along with the double gaussian solutions for the respective slices 
(A1 and C: solid lines; A2 and B: dashed lines).}
\end{figure}
\clearpage

\thispagestyle{empty}
\begin{figure}[h]
\vspace{7.0 truein}
\includegraphics{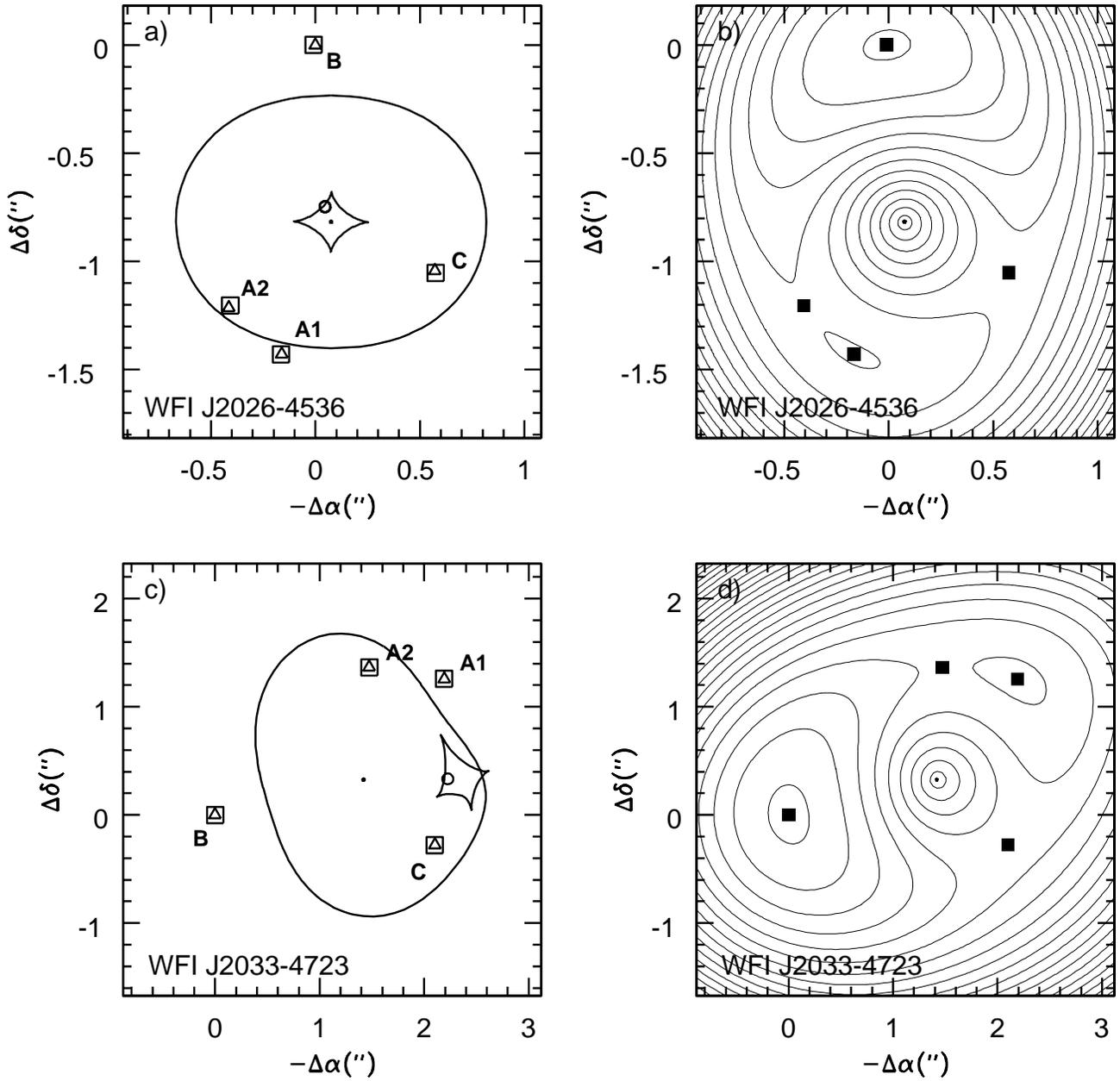}
\caption{{\it Panel a}: ISx+ model results for WFI~J2026$-$4536
showing the inner diamond caustic and outer critical line, the predicted image positions 
(squares), the observed image positions (triangles), and the model source position (open 
circle).  {\it Panel b}: Contour plot of WFI~J2026$-$4536 time delay function.
{\it Panels c and d}: Same as (a) and (b), but showing ISISx+ model results for 
WFI~J2033$-$4723.}
\end{figure}
\clearpage

\thispagestyle{empty}
\begin{figure}[h]
\vspace{7.0 truein}
\includegraphics{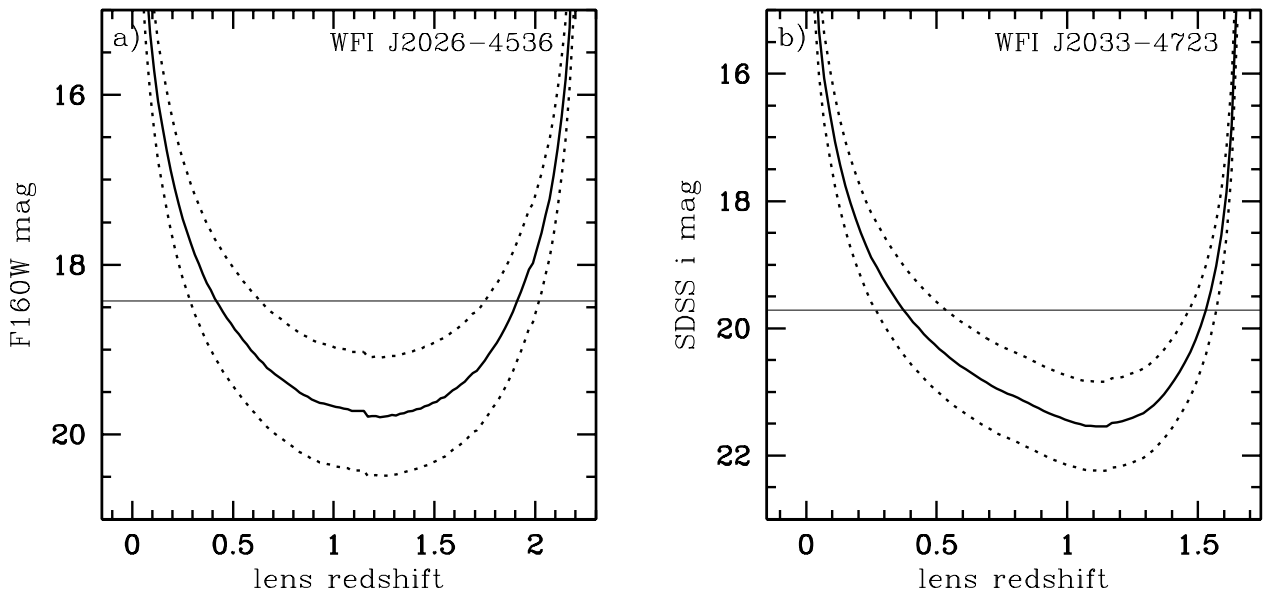}
\caption{Predicted lensing galaxy magnitudes for WFI~J2026$-$4536 
({\it panel a}) and WFI~J2033$-$4723 ({\it panel b}) as a function of the lens redshift 
(solid curve), along with $\pm 1\sigma$ errors (dashed curves) expected from the scatter 
in the Faber-Jackson relationship (Dressler et al. 1987).  The horizontal lines are the 
measured galaxy magnitudes.}
\end{figure}
\clearpage

\thispagestyle{empty}
\begin{figure}[h]
\vspace{7.0 truein}
\includegraphics{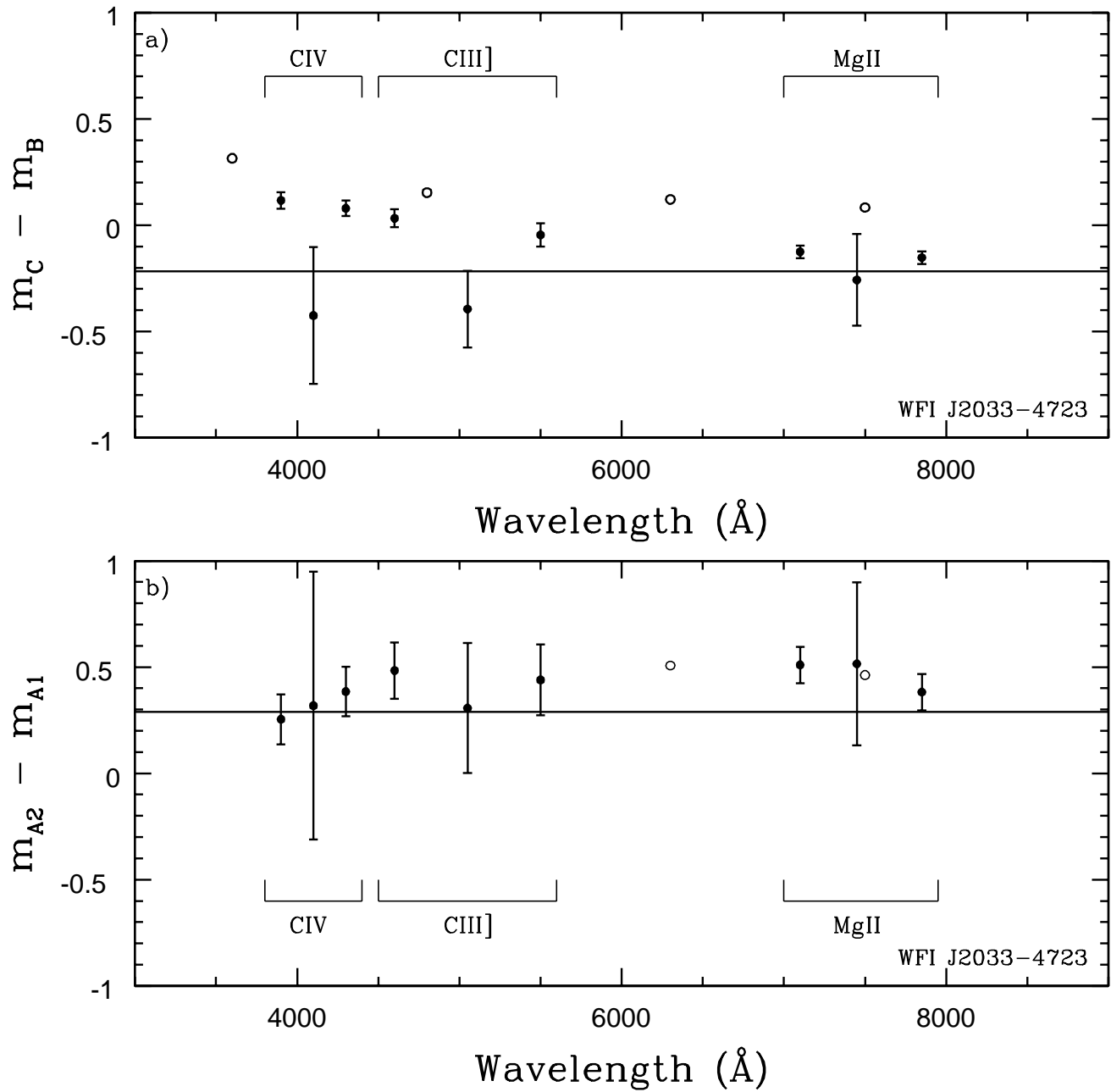}
\caption{{\it Panel a}: Magnitude differences for WFI~J2033$-$4723 
components B and C as a function of observed wavelength.  Solid points are from IMACS spectra
and are grouped in three's for each emission feature, consisting of the central emission line 
ratio and the two bracketing continuum ratios.  Open circles are the Magellan broadband 
$ugri$ flux ratios obtained one month earlier. {\it Panel b}: Same as (a), but for
WFI~J2033$-$4723 components A1 and A2.}
\end{figure}
\clearpage


\end{document}